\documentclass[12pt,titlepage,a4paper,draft]{report}

\usepackage{amsmath}
\usepackage{amssymb}
\usepackage{psfig}
\usepackage{epsfig}
\usepackage{latexsym}
\usepackage{graphicx}
\usepackage{epic}
\usepackage{eepic}
\usepackage{setspace}
\usepackage{supertabular}
\usepackage{psfrag}
\usepackage{rotating}
\usepackage{wrapfig}
\usepackage{fancyhdr}
\usepackage[dvips]{color}
\usepackage{cite}
\usepackage{makeidx}
\usepackage[english]{babel}
\psfull

\renewcommand{\chaptermark}[1]%
		{\markboth{#1}{}}
\renewcommand{\sectionmark}[1]%
		{\markright{\thesection\ #1}}
\makeindex

\begin{document}
\addtocontents{toc}{\protect\vspace{4ex}}
\pagenumbering{roman}

\newcommand{\beq}{\begin{equation}}
\newcommand{\eeq}{\end{equation}}

\newcommand{\beqa}{\begin{eqnarray}}
\newcommand{\eeqa}{\end{eqnarray}}

\newcommand{\bitem}{\begin{itemize}}
\newcommand{\eitem}{\end{itemize}}

\newcommand{\tabref}[1]{Tab.~\ref{#1}}
\newcommand{\Eqref}[1]{Eq.~(\ref{#1})}
\renewcommand{\eqref}[1]{(\ref{#1})}
\newcommand{\figref}[1]{Fig.~\ref{#1}}

\newcommand{\eg}{{\it e.g.}}
\newcommand{\ie}{{\it i.e.}}
\newcommand{\etal}{{\it et al.\/ }}

\newcommand{\mrm}{\mathrm}
\newcommand{\mbf}{\mathbf}
\newcommand{\doo}{\partial}
\newcommand{\bfr}{\mathbf{r}}
\newcommand{\bfp}{\mathbf{p}}
\newcommand{\bfv}{\mathbf{v}}

\newcommand{\Psihat}{\hat\Psi}
\newcommand{\psihat}{\hat\psi}
\newcommand{\Psihatdag}{\hat\Psi^\dag}
\newcommand{\psihatdag}{\hat\psi^\dag}
\newcommand{\Psihatr}{\hat\Psi(\textbf{r})}
\newcommand{\psihatr}{\hat\psi(\textbf{r})}
\newcommand{\Psihatdagr}{\hat\Psi^\dag(\textbf{r})}
\newcommand{\psihatdagr}{\hat\psi^\dag(\textbf{r})}
\newcommand{\firc}{\phi^*(\textbf{r})}
\newcommand{\fir}{\phi(\textbf{r})}
\newcommand{\psir}{\psi(\textbf{r})}

\newcommand{\Psihatdagrp}{\hat\Psi^\dag(\textbf{r}')}
\newcommand{\Psihatrp}{\hat\Psi(\textbf{r}')}

\newcommand{\bfrp}{\textbf{r}'}

\newcommand{\defas}{\mathrel{\raise.095ex\hbox{$:$}\mkern-4.2mu=}}

\newcommand{\nn}{\nonumber}
\newcommand{\half}{\frac{1}{2}}
\newcommand{\pfrac}[2]{\left( \frac{#1}{#2}\right)}
\newcommand{\lrp}[1]{\left( #1 \right)}
\newcommand{\lrpx}[1]{\left[ #1 \right]}
\newcommand{\ti}[1]{\tilde{#1}}

\newcommand{\viite}[4]{\bibitem{#1} #2, \emph{#3,} #4.}
\newcommand{\viitex}[3]{\bibitem{#1} #2, #3.}
\newcommand{\viitek}[4]{\bibitem{#1} #2, \emph{#3} #4.}

\renewcommand{\epsilon}{\varepsilon}
\newcommand{\comment}[1]{}
\setlength{\topmargin}{-1cm}
\begin{titlepage}
\vspace*{-30pt}
\mbox{}
\hrule
\vspace*{10pt}
\mbox{}\\
\begin{minipage}[t]{15mm} 
\vspace{-14pt}

\epsfig{file=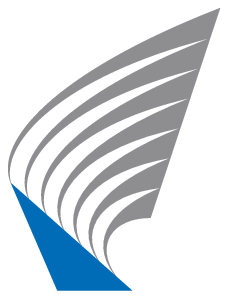, width=15mm}
\end{minipage}
\begin{minipage}[t]{200mm}
%\vspace{4mm}
\begin{tabbing}
Sopiva väli \=TEKNILLINEN KORKEAKOLU12345678901234567 \= 12345667890123456678 \= \kill

\>Helsinki University of Technology \\
\>Faculty of Information and Natural Sciences \\
\>Department of Applied Physics \\
%\> \today\\
\\

\end{tabbing}
\end{minipage}
\hrule
\begin{tabbing}
fdsa \= blaa blaa\= \kill
\\
\end{tabbing}

\begin{center}
\Large \bf Licentiate Thesis\\
\end{center}
\begin{tabbing}
\\
\end{tabbing}
\begin{center}
\LARGE \bf Entanglement-Enhanced Quantum Key Distribution \\
\end{center}
\begin{tabbing}
TEKNILLINEN KORKEAKOLU1234567890123456 \= ERIKOISTYO12\= \kill\\
\\
\\
\end{tabbing}
\begin{center}
\Large \bf Olli Ahonen \\
\end{center}
\begin{tabbing}
\\
\end{tabbing}
\begin{center}
\begin{tabular}{ll}
\large Supervisor:& \large Acad.\ Prof.\ Risto Nieminen\\
\parbox{0mm} {\vspace{2mm} } % \parbox{10mm} {\vspace{-1mm} } \\ 
\large Instructor:& \large Doc.\ Mikko M\"ott\"onen \\
\end{tabular}
\end{center}
\begin{tabbing}
\\
\end{tabbing}
\vspace*{15pt}
\hrule
\large Otaniemi\\
\large February 21, 2009
\end{titlepage}

\thispagestyle{empty}
{\textsc{Helsinki University of Technology\hfill{Abstract}}}
\vspace{1mm}

\fbox{\fbox{
\begin{minipage}{0.93\linewidth}

\begin{tabular}{p{33mm}l}
\bf{Author:}  				& Olli Ahonen  \\
\end{tabular}
{\vspace{-5mm}\center{\rule{0.99\linewidth}{1pt}}} \\
\begin{tabular}{p{33mm}l}
\bf{Faculty:}  					& Faculty of Information and Natural Sciences \\
\bf{Department:}				& Department of Applied Physics \\
\bf{Research field:}	  		& Engineering Physics, Theoretical and \\ & Computational Physics
\end{tabular}
{\vspace{-3mm}\center{\rule{0.99\linewidth}{1pt}}} \\
\begin{tabular}{p{33mm}l}
\bf{Title:}					& Entanglement-Enhanced Quantum Key Distribution\\
\bf{Title in Finnish:}	& Kietomalla kohennettu kvanttisalaus \\ 
\end{tabular}
{\vspace{-3mm}\center{\rule{0.99\linewidth}{1pt}}} \\
\begin{tabular}{p{33mm}l}
\bf{Chair:}	  				& Tfy-3 (Physics) \\
\bf{Supervisor:}			& Academy Professor Risto Nieminen\\
\bf{Instructor:}				& Docent Mikko M\"ott\"onen\\
\end{tabular}
{\vspace{-3mm}\center{\rule{0.99\linewidth}{1pt}}} \\
\begin{tabular}{p{13.5cm}}
\vspace{5pt}
Quantum key distribution (QKD) allows two spatially separated parties to securely generate a cryptographic key. The first QKD protocol, published by C.\ H.\ Bennett and G.\ Brassard in 1984 (BB84), describes how this is achieved by transmitting individual qubits and exchanging classical authenticated information. Any attempt to eavesdrop on the protocol introduces errors detectable by the legitimate parties.
\vspace{10pt}

This Licentiate Thesis studies the recently introduced EEQKD protocol which builds on BB84. In EEQKD, the qubits sent individually in BB84 are entangled and thus not directly available to an eavesdropper who is, in this protocol, provided only one-by-one access to the transmission. The maximal information an eavesdropper can gain using a straightforward intercept-resend (IR) attack, is obtained for a given error rate. The secure key generation rate of EEQKD is estimated in practical scenarios including qubit loss and quantum channel noise. In addition, an exquisite vulnerability is exposed: For a particular setting of qubit entanglement, paradoxically proving most useful in the face of an IR attack, EEQKD reduces to BB84.

\vspace{10pt}

Part of this research has been published in O.\ Ahonen et al., Physical Review A {\bf 78}, 032314 (2008).
\end{tabular}
{\vspace{-3mm}\center{\rule{0.99\linewidth}{1pt}}} \\
\begin{tabular}{p{33mm}l}
\bf{Keywords:}   & Quantum cryptography, Quantum key distribution, \\& BB84, Entanglement \\
\end{tabular}
\begin{tabular}{p{0.5\linewidth}l}
{\bf{Pages:}}  	 \hspace{20.5mm} 	  38 & {\bf{Date:}} February 21, 2009 \\
\end{tabular}
{\vspace{-5mm}\center{\rule{0.99\linewidth}{1pt}}} \\
\begin{tabular}{p{0.5\linewidth}l}
\bf{Approved:    }  	  & \bf{Location:   }

\end{tabular}

\end{minipage}}}

\thispagestyle{empty}
{\textsc{Teknillinen korkeakoulu\hfill{Tiivistelm\"a}}}
\vspace{1mm}

\fbox{\fbox{
\begin{minipage}{0.93\linewidth}

\begin{tabular}{p{3cm}l}
\bf{Tekij\"a:}	  		& Olli Ahonen  \\
\end{tabular}
{\vspace{-5mm}\center{\rule{0.99\linewidth}{1pt}}} \\
\begin{tabular}{p{3cm}l}
\bf{Tiedekunta:}		  	& Informaatio- ja luonnontieteiden tiedekunta \\
\bf{Laitos:}				& Teknillisen fysiikan laitos \\
\bf{Tutkimusala:}		& Teknillinen fysiikka, teoreettinen ja \\ & laskennallinen fysiikka 
\end{tabular}
{\vspace{-3mm}\center{\rule{0.99\linewidth}{1pt}}} \\
\begin{tabular}{p{3cm}l}
\bf{Otsikko:}			& Kietomalla kohennettu kvanttisalaus \\
\bf{Title:}				& Entanglement-Enhanced Quantum Key Distribution\\
\end{tabular}
{\vspace{-3mm}\center{\rule{0.99\linewidth}{1pt}}} \\
\begin{tabular}{p{3cm}l}
\bf{Professuuri:}	  	& Tfy-3 (Fysiikka) \\
\bf{Valvoja:}			& Akatemiaprofessori Risto Nieminen \\
\bf{Ohjaaja:}			& Dosentti Mikko M\"ott\"onen\\
\end{tabular}
{\vspace{-3mm}\center{\rule{0.99\linewidth}{1pt}}} \\
\begin{tabular}{p{13.5cm}}
\vspace{5pt}
C.\ H.\ Bennett ja G.\ Brassard julkaisivat vuonna 1984 BB84:ksi kut\-su\-tun menetelm\"an, jolla toisistaan et\"a\"all\"a olevat osapuolet voivat lu\-o\-da kryptografisen avaimen. Me\-ne\-tel\-m\"a hy\"odynt\"a\"a yksiqubittitilojen ominaisuuksia ja autentikoitua klassisten viestien vaihtoa. Kaikki salakuunteluyritykset voidaan havaita niiden v\"aist\"am\"att\"a aiheuttamien virheiden perusteella.
\vspace{10pt}

T\"ass\"a lisensiaatinty\"oss\"a tutkitaan hiljattain esitelty\"a EEQKD-kvant\-ti\-avai\-men\-jako\-me\-ne\-tel\-m\"a\"a. Sen sijaan, ett\"a yksitt\"aisi\"a hiukkasia l\"ahetett\"aisiin erikseen kuten BB84:ss\"a, ne ryhmitell\"a\"an ja kiedotaan erityisell\"a muunnoksella. Kietoutuneet hiukkaset l\"ahetet\"a\"an toiselle osapuolelle siten, ett\"a salakuuntelu rajoittuu yhteen hiukkaseen kerrallaan, joten se on perustavan\-laatuisesti rajoittu\-nutta.

\vspace{10pt}

Ty\"oss\"a lasketaan sieppaus-uudelleenl\"ahetys-hy\"ok\-k\"a\-yk\-sel\-l\"a (SUH) saatava maksimaalinen tieto avaimesta kvanttikanavassa aiheutetun h\"airi\"on funktiona. Salaisen avaimen muodostamisnopeus arvioidaan k\"ayt\"ann\"on sovellusalaa vastaavissa olosuhteissa, joissa kvanttikanavassa on sek\"a h\"avi\"ot\"a ett\"a kohinaa. Lis\"aksi n\"aytet\"a\"an, miten salakuuntelija voi murtaa avaimenjaon BB84-tasolle, kun k\"aytet\"a\"an er\"ast\"a kietovaa muunnosta, joka taas tarjoaa eniten suojaa SUH:ta vastaan.

\vspace{10pt}

Osa tutkimuksesta on julkaistu artikkelissa O.\ Ahonen et al., Physical Review A {\bf 78}, 032314 (2008).
\end{tabular}
{\vspace{-3mm}\center{\rule{0.99\linewidth}{1pt}}} \\
\begin{tabular}{p{3cm}l}
\bf{Avainsanat:}   & Kvanttikryptografia, Avaimenjako, BB84, \\& Kietoutuminen \\
\end{tabular}
\begin{tabular}{p{0.5\linewidth}l}
{\bf{Sivuja:}}  	 \hspace{16.7mm} 	38 & {\bf{P\"aiv\"ays:}} 21.~helmikuuta 2009\\
\end{tabular}
{\vspace{-5mm}\center{\rule{0.99\linewidth}{1pt}}} \\
\begin{tabular}{p{0.5\linewidth}l}
\bf{Hyv\"aksytty:    }  	  & \bf{Sijainti:    }

\end{tabular}

\end{minipage}}}

\addtolength{\textheight}{-5cm}	
\addtolength{\topmargin}{-2cm}	
\addcontentsline{toc}{chapter}{Acknowledgements}

\chapter*{Acknowledgements}

Although we part ways sooner than originally planned, I wish to express my gratitude to Docent Mikko M{\"o}tt{\"o}nen for his guidance and support during the research that culminates in this Licentiate Thesis. I would also like to thank Academy Professor Risto Nieminen for enabling all of this. Many thanks belong to the Quantum Computing and Devices group for the pleasant working atmosphere at TKK.

Nokia Corporation and The Finnish Cultural Foundation are acknowledged for financial support.
\\
\\
\\
\\
\\
\\
\\
\\
\\
\\
\\
\\
\\
\\
\\
\\
\\
\\
\emph{Helsinki, \today}
\\
\\
\\
Olli Ahonen

\addtolength{\topmargin}{+2cm}	
\addcontentsline{toc}{chapter}{Table of Contents}
\tableofcontents

\setlength{\baselineskip}{3.5ex}	
\pagestyle{fancyplain}

\chapter{Introduction}\label{ch:intro}

\pagenumbering{arabic}

\section{Quantum Cryptography}
Cryptography is about concealing the meaning of exchanged messages from unintended recipients. Today, strong methods are known for this task, such as the Advanced Encryption Standard \cite{stallings}. All cryptographic methods not based on assumptions about the difficulty of certain computational tasks require a cryptographic key, a relatively short random bit sequence known strictly only to the sender and recipient of the messages. The drawback of these robust encryption methods is the requirement of the pre-existing key.

Quantum cryptography, or synonymously quantum key distribution (QKD), addresses the difficulty of agreeing on a cryptographic key which fulfills the requirements of randomness and secrecy. In QKD, these requirements are met based on the laws of quantum physics. Notable conceptions in this field include entanglement, the no-cloning theorem, and the related effect of unavoidable disturbance of a quantum state in acquiring any information about it. Entanglement---the correlation of distinct quantum states regardless of their distance inexplicable by classical physics---is the main enabler of QKD in protocols such as that proposed by A.\ K.\ Ekert in 1991 (E91) \cite{e91}. Besides these so called entanglement-based protocols, the prepare-and-measure scheme is another viable approach to QKD \cite{mapra76}. The 1984 protocol by C.\ H.\ Bennet and G.\ Brassard (BB84) \cite{bb84} represents the latter type, and carries the title of the first QKD protocol. In BB84, the no-cloning theorem and the unavoidable disturbance of an unknown quantum state upon its measurement assert that the key generated by transmission and measurement of qubits, quantum bits, is secure after known classical post-processing steps. For a more extensive introduction to quantum cryptography, see \cite{dippa} and references therein.

This Licentiate Thesis studies the Entanglement-Enhanced Quantum Key Distribution (EEQKD) protocol first introduced by the author in Ref.~\cite{dippa} and further analyzed in Ref.~\cite{omapra}. The protocol is reviewed briefly in the next section. The main results of \cite{dippa} and \cite{omapra} are summarized in Ch.~\ref{ch:infoqber}. Chapter \ref{ch:keyrate} discusses the rate at which actual key bits are generated by the protocol in a practical application environment. Chapter \ref{ch:eeattack} presents an attack specifically designed for the EEQKD protocol, and shows that this attack reduces the EEQKD protocol, with certain parameter values, to the BB84 scheme. A preliminary version of this attack was introduced in Ref.~\cite{dippa}. Finally, this study and its implications are summarized in Ch.~\ref{ch:conclusions}.

\section{The Protocol}
\label{sec:protocol}
The EEQKD protocol is a prepare-and-measure-type scheme based on the BB84 protocol. Entanglement in the protocol was designed not to \emph{guarantee} the security, which is already achieved with the underlying BB84 mechanisms, but to decrease the information available to an attacker and thus increase the rate at which the key is generated. This Thesis adopts the conventional terminology of QKD: The initiator of the protocol is Alice, who wishes to establish a cryptographic key with Bob. The attacker, or eavesdropper, Eve attempts to gain as much information on the key as possible, without Alice or Bob detecting this.

\begin{figure}[hbtp]
\begin{center}
\includegraphics[width=0.65\textwidth]{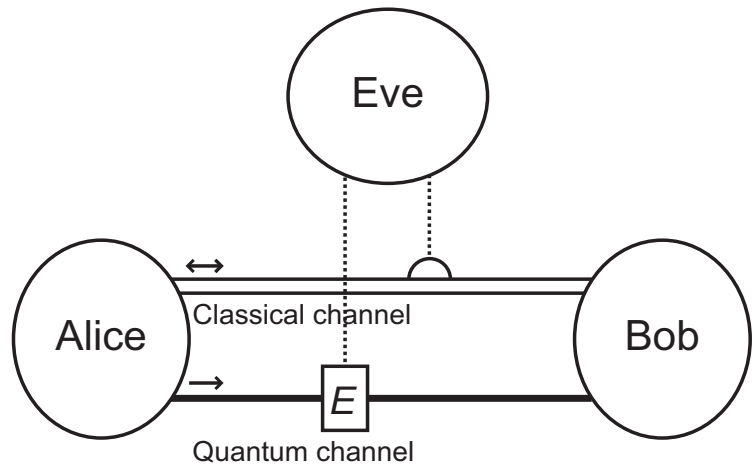}
\caption{General setup of the EEQKD protocol. Alice and Bob use a two-way classical channel and a one-way quantum channel. Eve has total control over the quantum channel, whereas she can only listen to the classical channel.}
\label{fig:bb84setup}
\end{center}
\end{figure}

Alice and Bob are connected by a quantum channel, i.e., Alice can send qubits to Bob. An ideal quantum channel conveys the qubit and its precise state. Eve is assumed to be in total control of the quantum channel. In addition, Alice and Bob have an authenticated classical channel allowing them to exchange classical information such that Eve can only read the messages. Figure \ref{fig:bb84setup} shows the general setup.

The EEQKD protocol is executed as follows. Alice prepares an $N$-bit string $a = a_1 a_2 \cdots a_N$, where each $a_i \in \{0,1\}$ is the outcome of the unbiased binary random variable $A_i$. String $a$ is the outcome of the composite $N$-bit random variable $A$. A large number of such strings forms what is known as Alice's raw key. Alice prepares a group of states $\{ |a_1;\alpha_1\rangle, |a_2;\alpha_2\rangle, \cdots, |a_N;\alpha_N\rangle \}$, where each $\alpha_i \in \{ z, x \}$ with uniform probability independently for each bit $i$. The states $|a_i;z\rangle$ and $|a_i;x\rangle$ are the distinct eigenstates of the Pauli matrices\footnote{The Pauli matrices are $\sigma_x = |0\rangle\langle1| + |1\rangle\langle0|$, $\sigma_y = -i|0\rangle\langle1| + i|1\rangle\langle0|$, and $\sigma_z = |0\rangle\langle0| - |1\rangle\langle1|$} $\sigma_z$ and $\sigma_x$, respectively. These $z$ and $x$ bases are denoted $\{|0\rangle, |1\rangle\}$ and $\{|+\rangle=(|0\rangle+|1\rangle)/\sqrt{2}, |-\rangle=(|0\rangle-|1\rangle)/\sqrt{2}\}$, respectively. Up to this point, the protocol is identical to BB84.

Alice applies an $N$-qubit transformation $U_N$ to the prepared qubit group. She then sends the qubits to Bob one by one, always delaying the transmission of the next qubit until acknowledgement from Bob of reception, sent via the authenticated classical channel. Bob can detect the reception of a qubit without disturbing its state by a quantum nondemolition (QND) measurement (see e.g.~\cite{pr-prl-92-190402}). Once Bob has received and stored the $N$ qubits, he applies $U_N^{\dagger}$ to the group. After this, the protocol again proceeds as BB84.

\begin{figure}[hbtp]
\begin{center}
\includegraphics[width=0.65\textwidth]{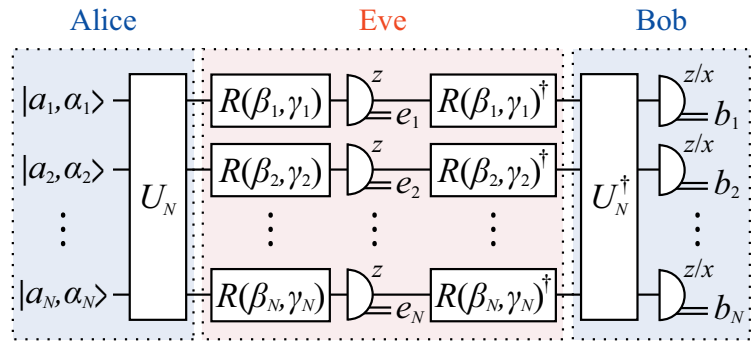}
\caption{Quantum circuit for the EEQKD protocol with an intercept-resend attack. This circuit generates $N$ raw key bits and is thus repeated as many times as needed. Bob's quantum nondemolition measurements, acknowledgement of qubit reception, and single-qubit memory are not shown. Semicircles represent projective measurements in the denoted bases. Gate $R(\beta,\gamma)$ is a single-qubit rotation with respect to axes $y$ and $z$, respectively. The $e_i$ are bits in Eve's view of Alice's raw key.}
\label{fig:circuit}
\end{center}
\end{figure}

Bob projectively measures each qubit in the eigenbasis of $\sigma_z$ or $\sigma_x$, chosen independently with uniform probability for each qubit. The measurement results $b_i \in \{0,1\}$ from a large number of groups form Bob's raw key. The described quantum transmission phase of the protocol is shown as a quantum circuit in Fig.~\ref{fig:circuit}. After the measurements, Alice and Bob compare their basis choices over the classical authenticated channel. They discard the bits in their raw keys corresponding to qubits transmitted and measured in different bases. The remaining bits, for which the transmission and measurement bases coincide, form the participants' sifted keys.

The sifted keys may still differ due to noise or eavesdropping in the quantum channel. Alice and Bob compare a small fraction of the keys, and based on the observed differences obtain an estimate of the quantum bit error rate (QBER), defined as the average probability of a bit flip in the sifted key. Classical error correction procedures (see e.g.~\cite{cascade}) can be applied if the QBER is less than 15\%. If eavesdropping is suspected, Alice and Bob employ privacy amplification, which allows them to reduce any eavesdropper's information to an arbitrarily low value while shortening the key. This results in the error-free and sufficiently secret final key. For a QBER in the range 15--25\% less efficient methods still exist for Alice and Bob to arrive at a final key \cite{rmpgisin}. For an alternative and more verbose description of the protocol model, see \cite{dippa}.

\section{Attacks}
In any attacks against a QKD protocol, the attacker aims at acquiring as much information on the key as possible, while inducing a minimal amount of errors detectable to Alice and Bob. The results presented in Chs.~\ref{ch:infoqber} and \ref{ch:keyrate} concern the EEQKD protocol under an intercept-resend (IR) attack. In the IR attack, Eve captures and projectively measures each individual qubit and sends Bob the state corresponding to her result. In BB84, Eve is able to guess correctly the basis used by Alice with probability $\frac{1}{2}$, in which case she obtains the correct key bit value. Otherwise she gains no information. During the comparison of transmission and measurement bases, Eve learns for each qubit if she used the correct basis or not. This leads to an average information of 0.5 bits on each sifted-key bit, and a QBER of 25\% due to the resent state being the correct one with probability 0.5 despite Eve's incorrect measurement basis. A detailed calculation is given in Ref.~\cite{dippa}. To induce a smaller QBER, Even can choose to interfere only with a fraction $\xi \in [0,1]$ of the transmitted qubits.

An incoherent attack is equivalent to Eve imperfectly cloning each transmitted qubit and measuring the clone in the correct basis after receiving the basis information from Alice \cite{rmp-scarani-1225}. Eve can adjust the fidelity of the clone state. The more information she gains on the key, the more the QBER increases. For a QBER below 25\%, the incoherent attack provides Eve slightly more information than the IR attack. Here, it is more beneficial for Eve to adjust the clone fidelity than to clone only a fraction of the qubits. Figure \ref{fig:qi} illustrates Eve's information as a function of the induced QBER for the attacks.

\section{Lower Bound on Security}
The security of the BB84 protocol has been proven against an eavesdropper with unlimited quantum computing capabilities \cite{inamoriproof}. This means that Alice and Bob can reliably estimate Eve's amount of knowledge on the generated key, and they can reduce this knowledge to an arbitrarily small value. The proof given in Ref.~\cite{inamoriproof} holds also for more restricted attacks such as IR and incoherent attacks. Furthermore, the proof assumes many realistic imperfections such as noise and loss in the quantum channel.

One may ask whether the addition of gates $U_N$ and $U_N^{\dagger}$ to the BB84 scheme could invalidate the security proof. From Fig.~\ref{fig:circuit}, it is immediate that in the worst case, the added gates are completely under Eve's control. But in this case, the circuit represents BB84 under an attack with some $N$-qubit quantum computation, and unconditional security for this setting has been proved. Therefore, EEQKD cannot be less secure than BB84, in any setting. What is more, calculations of the final bit rate, the rate at which bits are acquired for the final key, for BB84 also apply to the EEQKD protocol, assuming an ideal quantum channel. That is, in the ideal setting, EEQKD cannot perform worse than BB84. Noise and loss in the quantum channel can, however, change this as is shown in Ch.~\ref{ch:keyrate}.

\chapter{Information Gain of the Intercept-Resend Attack}
\label{ch:infoqber}

This Chapter studies the relationship between Eve's knowledge on the generated key and the QBER that Alice and Bob observe in the EEQKD protocol. Eve is assumed to employ the IR attack, and allowed to perform the included measurements in any basis. A definite upper bound to Eve's information is obtained for a given QBER. The calculations and results of this Chapter were published in Ref.~\cite{omapra}. Most of the calculations are laid out in more detail in Ref.~\cite{dippa}.

\section{Attacker's Information}
\label{sec:iae}
Let $e=e_1e_2\cdots{}e_N$, with each $e_i \in \{0,1\}$, denote the outcomes of Eve's measurements for the entangled group of $N$ qubits. The bit string $e$ is the outcome of the random variable $E$. Eve is allowed measurements in any basis, which is equivalent to allowing her arbitrary single-qubit gates and measurements in the $z$ basis. Any single-qubit gate can be decomposed to Bloch-sphere rotations about the axes $y$ and $z$ followed by a phase shift by $\phi$ as $e^{i\phi}R_z(\varphi) R_y(\beta) R_z(\gamma)$. To implement an arbitrary measurement basis, Eve performs a suitable rotation sequence prior to the measurement. Since Eve measures in the $z$ basis, the final rotation as well as the phase shift are irrelevant. To minimize the induced disturbance, Eve performs the opposite rotations after the measurement which can be assumed projective. Hence, Eve sends qubit $i$ in the state
\begin{equation}
|\Psi_i^{\mathrm{Eve}}(\beta, \gamma)\rangle = [R_y(\beta) R_z(\gamma)]^{\dagger} |e_i;z\rangle,
\end{equation}
to Bob. If Eve had somehow learned of another state leading to less errors at Bob's end, she would have obtained information on the key bit by some other means than the measurement. For individual unentangled qubits, this is not possible and resending $|\Psi_i^{\mathrm{Eve}}(\beta, \gamma)\rangle$ is the optimal strategy for Eve. In EEQKD, however, the optimal state may depend on Eve's earlier measurement results of the same group. We therefore state that this is a good IR attack strategy for Eve, but optimality cannot be claimed. This IR attack in EEQKD is shown in Fig.~\ref{fig:circuit}.

The random variables $A$ and $E$ allow us to precisely quantify Eve's information on the raw key. Since Eve learns which bits of the raw key contribute to the sifted key, her knowledge on the sifted key is directly described, as well. The per-bit mutual information of $A$ and $E$ is~\cite{nielsen}
\begin{equation}
I(A,E) = \frac{1}{N}[H(A) + H(E) - H(A,E)],
\label{eq:mutualinformation}
\end{equation}
where $H(\cdot)$ denotes the Shannon entropy and $H(\cdot,\cdot)$ the joint entropy. The mutual information or alternatively the entropies are averaged over Alice's basis choices $\alpha = \alpha_1\alpha_2\cdots\alpha_N$ which Eve learns during the sifting. The entropies are thus written as
\begin{equation}
H(E) = \frac{1}{2^{N}} \sum_{\alpha} H_{\alpha}(E) = -
\frac{1}{2^{N}} \sum_{\alpha,e} p(e|\alpha) \log_2 p(e|\alpha),
\end{equation}
and
\begin{equation}
H(A,E)  = \frac{1}{2^{N}} \sum_{\alpha} H_{\alpha}(A,E)
            = - \frac{1}{2^{N}} \sum_{\alpha,a,e} p(a,e|\alpha) \log_2
            p(a,e|\alpha).
\end{equation}
The entropy of $A$ is constant, $H(A) = N$.

\section{Quantum Bit Error Rate}
The quantum bit error rate is defined as the average probability of a different value of one bit, a bit flip, in Alice's and Bob's sifted keys. For each qubit of the $N$ qubits in a group, the QBER is
\begin{equation}
\mathrm{QBER}_j = \frac{1}{4} \sum_{\alpha_j = z}^x \sum_{a_j = 0}^1 p(B_j = \bar{a}_j | A_j = a_j; \alpha_j), \quad j = 1,\ldots,N,
\end{equation}
where $B_j$ is the random variable describing Bob's measurement of the $j$th qubit, and the bar denotes a bit flip, i.e., the logical {\sc not} operation. In the following, the QBER $q$ is the average of the single-qubit QBER's.

In practice, Alice and Bob cannot demand that they always observe a zero QBER, since various mechanisms of noise in the quantum channel typically lead to $q$ being in the range 1--10\% even without eavesdropping \cite{noise1,noise2,noise3,noise4}. Hence, Alice and Bob must tolerate a finite QBER. Only if $q$ is significantly larger than the estimated value for the equipment and the channel, can they be certain that the transmission is being eavesdropped. In the IR attack, this means that Eve adjusts the fraction $\xi$ suitable for inducing the QBER that Alice and Bob expect. Since it is quite possible for Eve to replace a noisy channel with a noiseless one and induce all of the observed QBER by eavesdropping, Alice and Bob must attribute all errors to Eve. For a given value of $q$, Eve's maximal information is thus determined by the maximum of the ratio $I(A,E)/q$.

\section{Two-Qubit Entanglement: Results}
\label{sec:2results}
This section shows Eve's maximal information for a given QBER in the case $N=2$, i.e., two-qubit entanglement. Gate $U_2$ is decomposed as
\begin{equation}
U_2 = C(\mathbf{c}) (k_{1,1} \otimes k_{1,2}) = \exp \left[ \frac{i}{2} \left(c_1\, \sigma_x \otimes \sigma_x + c_2\, \sigma_y \otimes \sigma_y + c_3\, \sigma_z \otimes \sigma_z\right) \right] (k_{1,1} \otimes k_{1,2}),
\end{equation}
where $\mathbf{c} = (c_1,c_2,c_3) \in [0,2\pi]^{\times 3}$ together with the arbitrary single-qubit gates $k_{1,l}$ para\-metrize the relevant two-qubit gates \cite{dippa}. In addition, we set $k_{1,1} = k_{1,2}$, relaxing of which can only decrease Eve's maximal information.

\begin{figure}[hbtp]
\begin{center}
\includegraphics[width=0.65\textwidth]{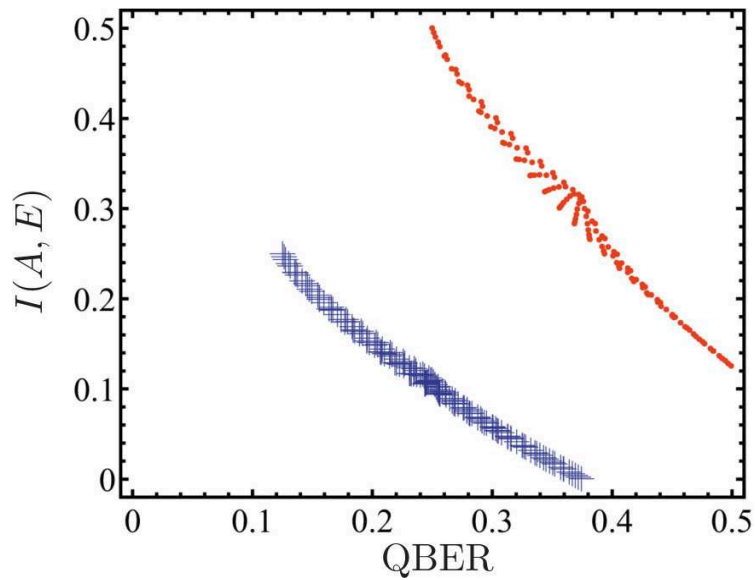}
\caption{Eve's information on the key and the induced QBER for different gates $U_2$. Eve uses the IR attack measuring in the $z$ basis. In the upper set, where both qubits are intercepted, the information is in the range 0.125--0.5. In the lower set, Eve intercepts only one of the qubits, and her information ranges from 0 to 0.25.}
\label{fig:csweep}
\end{center}
\end{figure}

Figure \ref{fig:csweep} shows a full sweep over the parameter space $\mathbf{c} \in [0,2\pi]^{\times 3}$ when Eve uses the IR attack in $z$ basis, without single-qubit gates. The upper set of points corresponds to Eve measuring both qubits, and the lower to Eve measuring only one of them. As one moves from the topmost to the undermost point of each set, the protocol shifts from BB84, where $U_2$ is the identity operator, to the maximally entangling EEQKD protocol with
\begin{equation}
U_2 = C\!\left(0,\frac{\pi}{2},0\right) = \left(I_1\otimes I_1 + i \sigma_y \otimes \sigma_y\right)/\sqrt{2},
\label{eq:u2star}
\end{equation}
where $I_1$ is the single-qubit identity operator. It is interesting to note that with this gate, Eve's information is strictly zero if she intercepts only one of the qubits. Furthermore, it remains zero for any measurement basis. Let $U_2^{\displaystyle{\star}}$ denote this gate. If Eve measures both qubits, one in basis $\sigma_z$ and the other in $\sigma_y$, her mutual information is 0.25, and the QBER is 0.375, which is optimal for Eve. Gate $U_2^{\displaystyle{\star}}$ is generalized to $N$ qubits in Sec.~\ref{sec:ustar}.

\begin{figure}[hbtp]
\begin{center}
\includegraphics[width=0.65\textwidth]{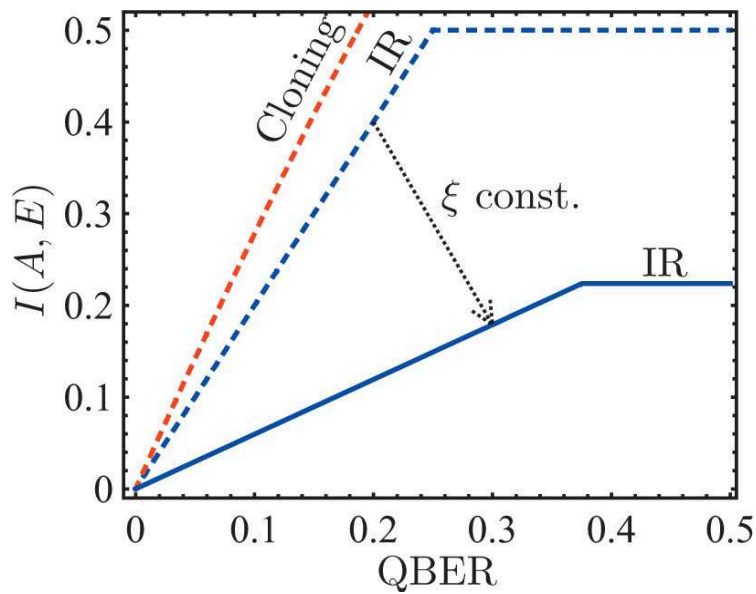}
\caption{Eve's maximal information on the key as a function of the QBER Alice and Bob observe. Eve uses the IR attack choosing measurement bases freely (blue lines) or the cloning attack (red dashed line). The dashed lines correspond to BB84 and the solid line to the EEQKD protocol with the optimal gate $U_2 = C\left(\frac{\pi}{32},\frac{3\pi}{8},\frac{\pi}{32}\right)$. The arrow shows how the setting changes from BB84 when the fraction of intercepted qubits $\xi$ is held constant. For a QBER below 25\%, Eve's information is decreased by more than 70\%.}
\label{fig:qi}
\end{center}
\end{figure}

In the following, Eve is allowed to choose her measurement bases freely. To obtain the optimal $U_2$, one needs to solve the twofold optimization problem of Alice and Bob minimizing Eve's maximal information for a given QBER. In other words, the problem is to find $\min_{\mathbf{c}} \max_{\{\beta_1,\gamma_1,\beta_2,\gamma_2\}} \left[ I(A,E)/q \right]$ and the optimizing parameter values. There are several parameter values solving the problem. One of the optimal values for Alice and Bob is
\begin{equation}
\mathbf{c}^* = \left(\frac{\pi}{32},\frac{3\pi}{8},\frac{\pi}{32}\right).
\end{equation}
This yields $I(A,E) \approx 0.2237$ and $q = 0.375$, if $\xi=1$ and if Eve uses her corresponding optimal values $\left(\beta_1,\gamma_1,\beta_2,\gamma_2\right) = \left(\frac{\pi}{8},0,\frac{\pi}{2},\frac{\pi}{2}\right)$. Figure \ref{fig:qi} shows Eve's maximal information as a function of the observed QBER given $U_2 = C(\mathbf{c}^*)$. The information provided by an IR attack in the BB84 protocol is also shown, its explicit formula is $I(A,E) = 2q$.

\section{Gate Construction for $N$ Qubits}
\label{sec:ustar}
Let us generalize the gate $U_2^{\displaystyle{\star}}$, which produces nearly optimal results for Alice and Bob in the two-qubit case, to an arbitrary number of entangled qubits $N$. Let $U_N^{\displaystyle{\star}}$ denote this generalized gate. There are two equivalent choices of different parity for the gate for each value of $N$. We denote them $U_{N,\mathrm{even}}^{\displaystyle{\star}}$ and $U_{N,\mathrm{odd}}^{\displaystyle{\star}}$, both of which produce the same maximal information for Eve. We define $U_{1,\mathrm{even}}^{\displaystyle{\star}} = I_1$ and $U_{1,\mathrm{odd}}^{\displaystyle{\star}} = \sigma_y$. The $(N+1)$-qubit gate is obtained with the recursive formula
\begin{equation}
U_{N+1}^{\displaystyle{\star}} = \frac{1}{\sqrt{2}} \left[I_1 \otimes U_N^{\displaystyle{\star}} \pm i \sigma_y \otimes
\left(P_N U_N^{\displaystyle{\star}}\right)\right],
\label{stargate}
\end{equation}
where $P_N = \sigma_{y} \otimes I_1^{\otimes N-1}$ if $N\geq2$ and $P_1 = \sigma_y$. Either of the two signs can be chosen independently for each step. For instance, the gate given in Eq.~(\ref{eq:u2star}) is of even parity, $U_{2,\mathrm{even}}^{\displaystyle{\star}}$, and produced by the application of the upper sign in the formula. The gate $U_N^{\displaystyle{\star}}$ restricts the information gained by any IR attack to at most $1/(2N)$. A proof of this is given in the Appendix in Ref.~\cite{omapra}.

\chapter{Key Generation Rate}
\label{ch:keyrate}

This chapter studies the rate at which the EEQKD protocol allows Alice and Bob to generate bits for the final key. This final bit rate is arguably the most important figure for a QKD protocol once unconditional security has been proven \cite{rmpgisin}. After all, in a practical setting, one would prefer to generate the key as fast as possible or, alternatively, use the least necessary technological resources to achieve the required key generation rate. Throughout this chapter, Eve is assumed to use the IR attack described in Sec.~\ref{sec:iae}.

\section{Relative Key Rate}
Since the actual final bit rate $R_{\mathrm{net}}$, measured in bits per second, heavily depends on the equipment chosen for the realization of the protocol, we use the relative key rate $r=R_{\mathrm{net}}/R_{\mathrm{sift}}$, where $R_{\mathrm{sift}}$ is the rate at which sifted key bits are generated. The relative key rate thus reflects the capability of a protocol to deliver secret, error-free bits, independent of the used technology. It allows for the comparison of QKD protocols without regard to any specific implementation.

The exact relative key rate depends on the error correction and privacy amplification procedures applied to the sifted key. It is also a function of the amount of errors and their possible correlations in the key, as well as of the amount and structure of the information that Eve may have on the key \cite{bouwmeester}. Therefore, calculation of the exact value of $r$ requires detailed knowledge and assumptions of the protocol and the participants' actions during its execution.

Since we do not wish to fix every smallest detail of the entire key distribution procedure, we take the information-theoretic approach \cite{renner-itt} to obtain an estimate of the relative key rate $r$. In this paragraph, we follow Ref.~\cite{rmpgisin}. The relative key rate is determined by the difference of Bob's and Eve's knowledge of Alice's sifted key as
\begin{equation}
r = R_{\mathrm{net}}/R_{\mathrm{sift}} = I(A,B) - I(A,E).
\label{r-infos}
\end{equation}
A non-positive key rate given by this formula might, in some cases, be circumvented by advanced techniques. Equation (\ref{r-infos}), however, suffices for our purposes. Errors in the quantum channel decrease the correlation of Bob's and Alice's sifted keys. Thus,
\begin{equation}
I(A,B) = 1 - H_{\mathrm{bin}}(q),
\end{equation}
where the binary entropy
\begin{equation}
H_{\mathrm{bin}}(q) = -q \log_2 q - (1-q)\log_2 (1-q),
\end{equation}
and finally,
\begin{equation}
r(q) = 1 - I(A,E) - H_{\mathrm{bin}}(q).
\end{equation}

In the BB84 scheme, the relative key rate with an IR attack is
\begin{equation}
r_{\mathrm{BB84}}(q) = 1 - 2q - H_{\mathrm{bin}}(q).
\label{eq:ratebb84}
\end{equation}
For the EEQKD protocol, Eve's information provided by an IR attack as a function of the QBER $q$ is determined by the slope $s$ of the line shown in Fig.~\ref{fig:qi}, i.e., $s = \max \left[ I(A,E)/q \right]$, where the maximization is over Eve's parameters for a given gate $U_N$. We denote the observed QBER in EEQKD by $q_E = \delta q$, and let $q$ denote the QBER which the innocent noise of the quantum channel induces with BB84. Below, it is demonstrated that the coefficient $\delta \geq 1$, i.e., entangling successive qubits for transmission can only increase the innocent error rate. The relative key rate is
\begin{equation}
r_{U_N}(q) = 1 - s q_E - H_{\mathrm{bin}}(q_E),
\label{eq:relkeyrate}
\end{equation}
because Alice and Bob must assume that all errors are due to Eve.

It is convenient to view the effects of different gates $U_2$ with reference to a fixed value of $q$. For this purpose, we define $q_{\mathrm{ref}} = 6\%$, which is a typical value for the QBER in a BB84-type scheme \cite{noise1,noise2,noise3,noise4}. This value yields a relative key rate of $r_{\mathrm{BB84}}(q_{\mathrm{ref}}) = 0.553$.

\begin{figure}[hbtp]
\begin{center}
\includegraphics[width=0.72\textwidth]{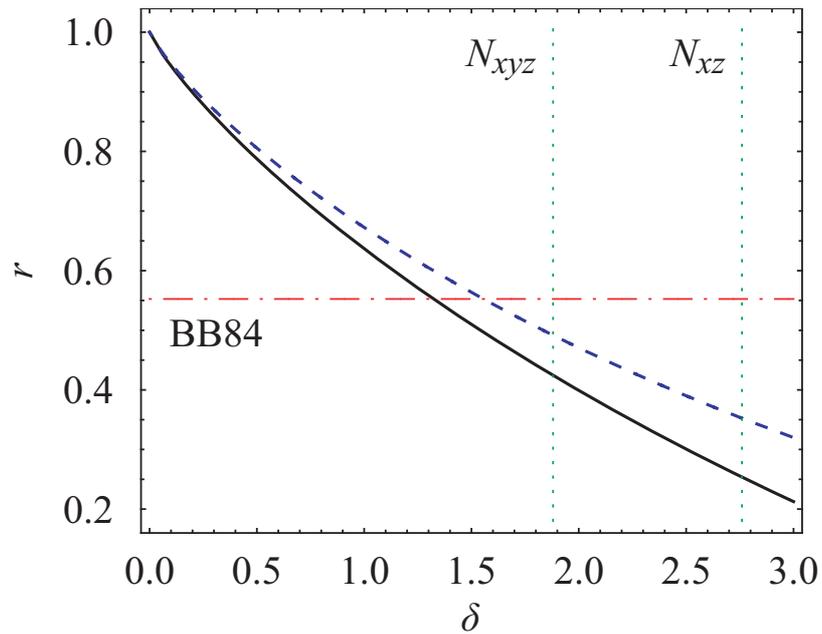}
\caption{Relative key rate as a function of the coefficient $\delta$ by which the use of an entangling gate changes the observed QBER. The QBER in BB84 is fixed to $q_{\mathrm{ref}} = 6\%$. The EEQKD protocol is shown in solid line for $U_2 = C(\mathbf{c}^*)$. The blue dashed line shows the maximal rate for the EEQKD protocol, corresponding to $s=0$. The relative key rate of the BB84 protocol is 0.553, shown in red dash-dotted line. The vertical green dotted lines mark $\delta$ for the gate $U_2^{\displaystyle{\star}}$ with noise operators $N_{xyz}$ and $N_{xz}$.}
\label{fig:r-delta}
\end{center}
\end{figure}

Figure \ref{fig:r-delta} shows the relative key rate for the optimized gate $U_2 = C\left(\frac{\pi}{32},\frac{3\pi}{8},\frac{\pi}{32}\right)$ which yields $s= 0.5965$ (see Ch.~\ref{ch:infoqber}). For comparison, the dashed blue line shows the rate for a protocol with $s=0$. This ideal protocol demonstrates how much the general structure of the EEQKD protocol could possibly improve $r$ over the BB84 scheme. It provides Eve with no information in the quantum transmission phase, that is, its line would, in Fig.~\ref{fig:qi}, lie on the horizontal axis. The key rate of the ideal protocol is equal to that of BB84 at $\delta = 1.555$. Therefore, any gate $U_N$ that leads to $\delta > 1.555$ will inevitably perform worse than the plain BB84 scheme, unless additional analysis and processing is applied. For the optimized protocol with $U_2 = C\left(\frac{\pi}{32},\frac{3\pi}{8},\frac{\pi}{32}\right)$, the condition for better performance is $\delta < 1.323$. The QBER is fixed at $q_{\mathrm{ref}}$.

\section{Quantum Channel Imperfections}
\label{sec:chanimperf}

The realization of a quantum channel for QKD has two main alternatives: optical fiber and the Earth's atmosphere. The creation of a qubit in a suitable state and its detection at Bob's end can be implemented in various ways in both alternatives. Most qubit creation methods employ either an attenuated laser or parametric downconversion. Avalanche photodiodes are the most common choice for detecting the qubit state. \cite{mauerer-recent}

Both fiber and atmosphere have two major drawbacks for QKD protocols: The quantum channel may lose a qubit altogether, or the state of the qubit may be altered such that Bob receives some other state than which Alice sent. These processes form the innocent noise of the quantum channel---noise present even without Eve. The process of qubit loss is also known as damping or attenuation. In fiber and in good atmospheric conditions, the damping is approximately --0.21 dB/km. Poor conditions in the atmosphere can raise the damping to a devastating --20 dB/km \cite{mauerer-recent}. Change of the qubit state during transmission contributes to the QBER, for which typical values range from 1\% to 10\% \cite{noise1,noise2,noise3,noise4}.

Let us quote the damping values reported in a recent QKD experiment carried out over a distance of 144 km in the atmosphere \cite{noise4}. The total damping for single photons in the channel was typically --30 dB, which means that only 0.1\% of photons transmitted reached the detector at Bob's end. The atmosphere contributed between --8 and --12 dB, photons not hitting the receiver between --10 and --16 dB, and optics --2 dB in total. The atmospheric damping is thus --0.07 dB/km. For instance, with the setup of \cite{noise4}, a distance of 10 km would lead to a damping of --(0.7 dB + 2 dB) $\approx$ --3 dB, which means that only half of the qubits would reach Bob's detectors, on average.

\section{Effect of Noise}
\label{sec:noiseeffect}
To obtain an estimate for the coefficient $\delta$ by which the EEQKD protocol increases the QBER relative to BB84, we assume that the quantum channel acts independently on each transmitted qubit with one of the two unitary noise operators
\begin{equation}
N_{xyz}(n) = [(1-n) I_1 + n(\sigma_x + \sigma_y + \sigma_z)]/\sqrt{4n^2-2n+1},
\label{eq:noisexyz}
\end{equation}
and
\begin{equation}
N_{xz}(n) = [(1-n) I_1 + n(\sigma_x + \sigma_z)]/\sqrt{3n^2-2n+1},
\end{equation}
where $n$ is the amplitude of the noise.

With the noise operator $N_{xyz}(n)$, we obtain a mapping between the QBER in the BB84 scheme $q$ and the amplitude $n$ through numerical calculation. The mapping is shown in Table \ref{map-nq}. Once the amplitude $n$ is known for each value of $q$, it is straightforward to calculate the corresponding QBER in EEQKD, $q_E = \delta q$. Note that $\delta$ and thus $q_E$ depend on the gate $U_N$ chosen by Alice and Bob.

\begin{table}[hbt]
\begin{center}
\caption{Mapping between the amplitude $n$ of the quantum channel noise operator $N_{xyz}(n)$ and the QBER the operator induces in the BB84 scheme.}
\label{map-nq}
\vspace{2mm}
\begin{tabular}[c]{lcccccccc}
\hline
\hline
$q$	& 2\%		& 4\%		& 6\%		& 8\% 		& 10\%		& 12\%		& 25\%			& 50\%		\\
$n$	& 0.0922	& 0.1274	& 0.1537	& 0.1758	& 0.1952	& 0.2129	& 0.3091		& 0.5000	\\
\hline
\hline
\end{tabular}
\end{center}
\end{table}

\subsection{Gate $U_2^{\displaystyle{\star}}$}
Let us study the gate $U_2^{\displaystyle{\star}}$, defined in Sec.~\ref{sec:2results}, under the above noise operators in more detail. Numerical calculation shows that the coefficient $\delta$ is exactly $2(1 - q)$ and thus the resulting QBER due to noise is
\begin{equation}
q_E^{\mathrm{noise}} = 2(q-q^2),
\end{equation}
provided that the noise operator is $N_{xyz}(n)$. For the noise operator $N_{xz}(n)$, the calculation yields an even higher QBER, namely
\begin{equation}
q_E^{\mathrm{noise}} = 3q - 4q^2.
\end{equation}
For the highest sensible error rate value $q=0.5$, both operators yield $q_E^{\mathrm{noise}} = 0.5$, as is expected. If $q = q_{\mathrm{ref}}$, the coefficient is $\delta = 1.88$ for the operator $N_{xyz}(n)$ and $\delta = 2.76$ for $N_{xz}(n)$. These values are shown in Fig.~\ref{fig:r-delta}, which demonstrates that the relative key rate is lower than that of BB84, if $q = q_{\mathrm{ref}}$.

\begin{figure}[hbtp]
\begin{center}
\includegraphics[width=0.72\textwidth]{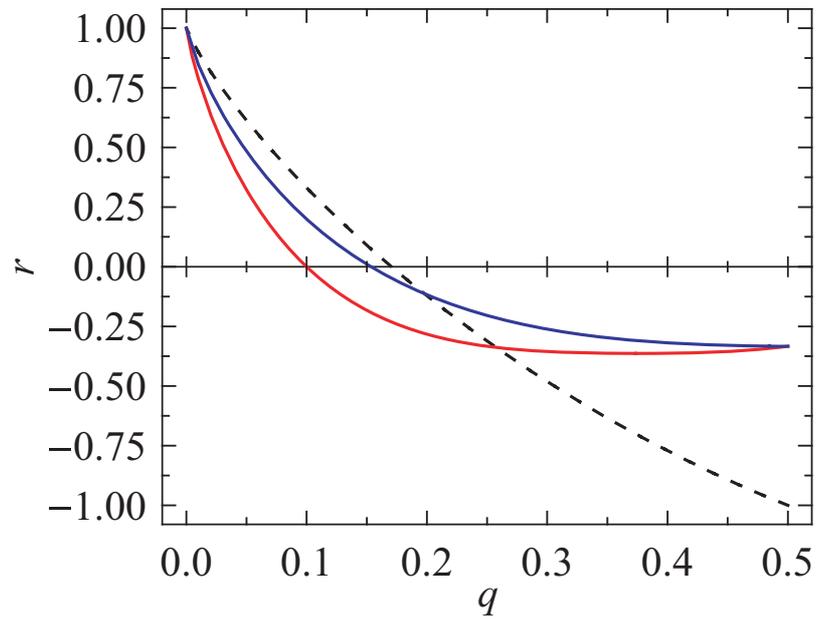}
\caption{Relative key rate under quantum channel noise for the gate $U_2^{\displaystyle{\star}}$ as a function of the QBER in BB84. The rate for the BB84 scheme is shown in dashed line, it is positive for $q < 17.05\%$. The upper blue line corresponds to the noise being due to operator $N_{xyz}$, and the lower red line due to $N_{xz}$. The rates are positive for $q < 15.36\%$ and $q < 10.00\%$, respectively.}
\label{fig:u2star-n-rates}
\end{center}
\end{figure}

The relative key rate is
\begin{equation}
r_{U_2^{\displaystyle{\star}}}(q) = 1 - \frac{2}{3} q_E - H_{\mathrm{bin}}(q_E),
\end{equation}
since the maximized slope is $s=\frac{0.25}{0.375}=\frac{2}{3}$. The rate is plotted in Fig.~\ref{fig:u2star-n-rates} as a funtion of $q$ for the two defined noise operators. The BB84 scheme, shown in dashed line in Fig.~\ref{fig:u2star-n-rates}, yields a positive key rate up to $q = 17.05\%$. Gate $U_2^{\displaystyle{\star}}$ yields a positive rate if $q < 15.36\%$ or $q < 10.00\%$ for quantum channel noise given by $N_{xyz}(n)$ or $N_{xz}(n)$, respectively. The rate is, however, lower than that of BB84.

A positive key rate is possible with BB84 up to $q = 25\%$, in which case more complex methods of quantum privacy amplifiction or classical advantage distillation must be used \cite{rmpgisin}. Whether the same holds for gate $U_2^{\displaystyle{\star}}$, or for the EEQKD protocol in general, remains to be studied.

If qubit loss remains at a tolerable level, it is possible that the EEQKD protocol with gate $U_2^{\displaystyle{\star}}$ performs better than the BB84 scheme at large error rates. Unfortunately, the practical situation for QKD at present is the opposite: Qubit loss is very high, of the order of 0.999, and the error rate is quite small.

\section{Effect of Qubit Loss}
In the BB84 scheme, the effect of qubit loss is to simply reduce the amount of bits in the sifted key by one for every photon lost in the quantum channel. In EEQKD however, the loss of one qubit in general affects the other entangled $N-1$ qubits, as well. Thus, qubit loss increases the QBER of the remaining qubits.

This section studies the effect of qubit loss in the case where exactly half of the transmitted qubits are lost in the quantum channel. This roughly corresponds to a transmission distance of 10km, as discussed in Sec.~\ref{sec:chanimperf}. We also make the reasonable assumption that the probability of loss, 0.5, is uniform and independent for each qubit. For simplicity, we set $N=2$, i.e., study two-qubit entanglement.

If qubit loss is 50\%, the probability for a qubit pair, entangled or not, to lose both of its qubits is $0.5\times0.5 = 0.25$. Symmetrically, the probability that neither qubit is lost is 0.25. The probability to lose exactly one qubit is 0.5. Entanglement between the qubits does not affect loss, as is demonstrated by the expected number of lost qubits: $2\times0.25 + 1\times0.5 = 1$ for a pair of qubits.

Entanglement does, however, affect the change of the state of the remaining qubit if exactly one qubit is lost. Let $q_{\mathrm{loss}}$ denote the QBER of the remaining qubit due to loss of the other. Let us optimistically assume that quantum channel noise induces the same QBER as in BB84, denoted $q$, for the remaining qubit. The total QBER of the remaining qubit is obtained as the exclusive-or of the two probabilities $q_{\mathrm{loss}}$ and $q$. It is
\begin{equation}
q_{\mathrm{loss}}^{\mathrm{tot}} = q + q_{\mathrm{loss}} - 2 q q_{\mathrm{loss}}.
\end{equation}
That is, a bit flip is due to noise or loss of the other qubit, but not both (since then the bit value would be correct). Only half of the transmitted qubit pairs lose exactly one qubit. The qubit pairs transmitted without loss contribute two bits per pair to the raw key. Their QBER $q_E^{\mathrm{noise}}$ may differ from $q$. The overall average QBER is thus
\begin{equation}
q_E = (q_E^{\mathrm{noise}} + q + q_{\mathrm{loss}} - 2 q q_{\mathrm{loss}})/2.
\label{eq:qtot}
\end{equation}
To simulate the effect of qubit loss, Bob may be assumed to replace each lost qubit with a constant state, e.g., $|0\rangle$.

\begin{figure}[hbtp]
\begin{center}
\includegraphics[width=0.72\textwidth]{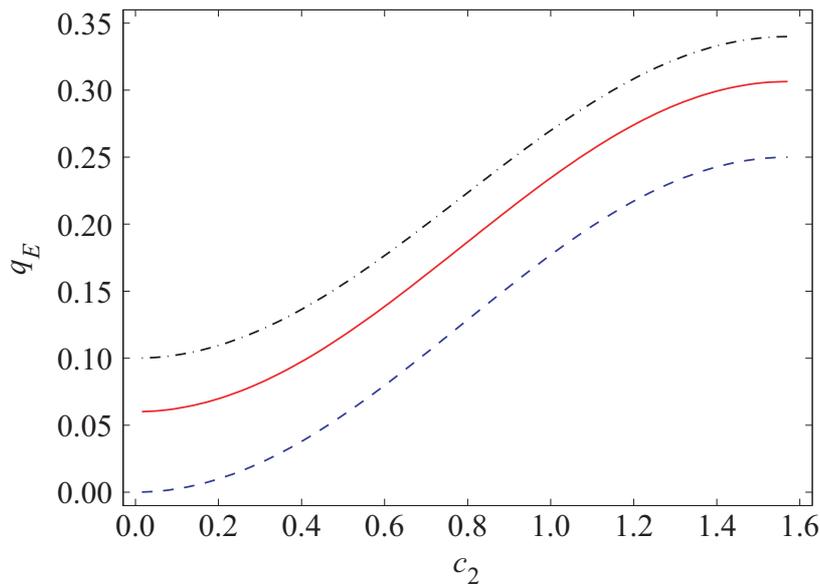}
\caption{Total average QBER as a function of the gate parameter $c_2$ for 50\% qubit loss. The other gate parameters $c_1 = c_3 = 0$. The QBER in BB84 is zero for the blue dashed line, 6\% for the red solid line, and 10\% for the black dash-dotted line. At $c_2 = 0$, the protocol is equivalent to BB84, and at $c_2 = \pi/2 \approx 1.57$, the gate is $U_2^{\displaystyle{\star}}$. }
\label{fig:qtot-c2}
\end{center}
\end{figure}

Figure \ref{fig:qtot-c2} shows $q_E$ for different values of the gate parameter $c_2$. This specific parameter was chosen because varying it between 0 and $\frac{\pi}{2}$ covers most of the relevant gates $U_N$ and corresponds to a shift from the BB84 scheme to using gate $U_2^{\displaystyle{\star}}$. The other two parameters are fixed to $c_1 = c_3 = 0$. Noise is given by Eq.~(\ref{eq:noisexyz}). For $q=0$, shown in dashed line, the overall average QBER increases from 0 to 25\%. For $q=q_{\mathrm{ref}}$, shown in solid line, $q_E$ increases from $q_{\mathrm{ref}}$ to 30.64\%. The dash-dotted line corresponds to $q=10\%$, and yields a $q_E$ between 10\% and 34.00\%. With $U_2^{\displaystyle{\star}}$, the effect of qubit loss is devastating: The remaining qubit of a pair of which the other qubit is lost, is completely scrambled, i.e., $q_{\mathrm{loss}} = 0.5$. For the optimized gate $C(\mathbf{c}^*)$ defined in Sec.~\ref{sec:2results}, $q_E = 21.37\%$ for $q=0$ and $q_E = 26.86\%$ for $q=6\%$. To conclude, the error rate induced by qubit loss can be very high and can thus decrease the key rate in EEQKD significantly, in addition to decrease due to quantum channel noise.

\section{Optimizing the Key Rate}

\subsection{Rate without Qubit Loss}
\label{sec:optimizenoloss}
We maximize the two-qubit relative key rate $r_{U_2}(q)$ for various values of the BB84 QBER $q$. The optimization is performed over the entangling gate parameters $\mathbf{c}$. The slope $s$ is the maximum obtained by optimization over Eve's parameters. Quantum channel noise, given by Eq.~(\ref{eq:noisexyz}), is taken into account with the approach described in Sec.~\ref{sec:noiseeffect}. We assume no loss of qubits in the quantum channel.

In addition to random initial values of $\mathbf{c}$ and random identical single-qubit gates before gate $C(c)$, we start the optimization at $U_2 = I_2$ (the BB84 scheme), at $U_2 = U_2^{\displaystyle{\star}}$, and at $\mathbf{c} = \mathbf{c}^*$. We iterate the optimization for this array of initial values over the $q$ values of 2\%, 4\%, 6\%, 8\%, 10\%, 12\%, and 25\%.

\begin{figure}[hbtp]
\begin{center}
\includegraphics[width=0.72\textwidth]{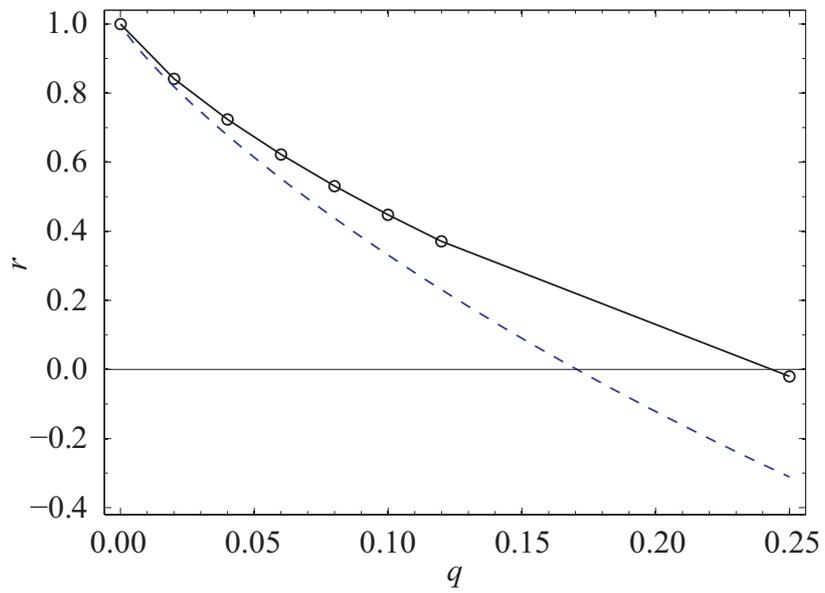}
\caption{Relative key rate as a function of the QBER in BB84. Quantum channel noise is given by the operator $N_{xyz}(n)$, and no qubit loss is assumed. The circles denote the rate for the optimal gate $C \left( \frac{\pi}{4},\frac{\pi}{4},\frac{\pi}{4} \right)$ at the discrete values of QBER for which the rate was maximized, and the solid line is drawn to guide the eye. The BB84 scheme is shown in blue dashed line.}
\label{fig:piper4}
\end{center}
\end{figure}

For all the employed values of $q$, gate $C \left( \frac{\pi}{4},\frac{\pi}{4},\frac{\pi}{4} \right)$ yields the highest relative key rate. Figure \ref{fig:piper4} shows the rate for this gate for the various values of $q$, denoted with circles, and for BB84 in blue dashed line. For $q = 25\%$, the optimal gate yields a key rate of $-0.02$. The negativity of the rate is a symptom of the fact that more advanced post-processing methods are required, even in the plain BB84 scheme. This setting is, however, irrelevant in the practical situation, where the QBER is significantly lower.

\subsection{Rate with Qubit Loss}
\label{sec:optimizeloss}
We repeat the optimization procedures described in the previous section for a lossy quantum channel. We use Eq.~(\ref{eq:qtot}), i.e., take into account quantum channel noise, given by Eq.~(\ref{eq:noisexyz}), and qubit loss, which is fixed to 50\%.

For QBER's in the range 2-12\%, the optimization did not reveal any gates which would yield a higher key rate than BB84; successful maximizations terminated at the gate being the identity operator. For $q = 25\%$, we were able to find a gate yielding a key rate of $-0.16$, which is higher than that for BB84. Again, the negativity of the rate shows that more advanced post-processing methods are required for large values of QBER.

\subsection{Correlated Bit Errors}
\label{sec:errcorr}
For highly entangling gates, the bit errors of an entangled pair are expected to be correlated. Let us optimistically assume that the errors are fully correlated. Then, the amount of bits exchanged, and thus disclosed to Eve, in the error correction phase is halved. This is reflected in the relative key rate as
\begin{equation}
r_{U_2}^{\mathrm{corr}}(q) = 1 - s q_E - \frac{1}{2} H_{\mathrm{bin}}(q_E),
\label{eq:corrrate}
\end{equation}
since Eve's information is now $s q_E + \frac{1}{2} H_{\mathrm{bin}}(q_E)$, the sum of information obtained via eavesdropping and by listening to the classical error correction procedure \cite{bouwmeester}.

We repeat the optimization procedures described in Sec.~\ref{sec:optimizenoloss} with the assumption of a 50\% qubit loss and using the key rate given by Eq.~(\ref{eq:corrrate}). The end result is the same as in Sec.~\ref{sec:optimizeloss}: For a QBER in the range 2-12\%, BB84 yields the highest key rate, and for $q = 25\%$, there is a non-identity gate yielding the highest rate. The latter is achieved approximately at $\mathbf{c} = (0.11,0.89,0.10)$, with a relative key rate of 0.30. This gate, however, gives no particular justification to assume full correlation of errors in the entangled qubit pair.

\begin{figure}[hbtp]
\begin{center}
\includegraphics[width=0.72\textwidth]{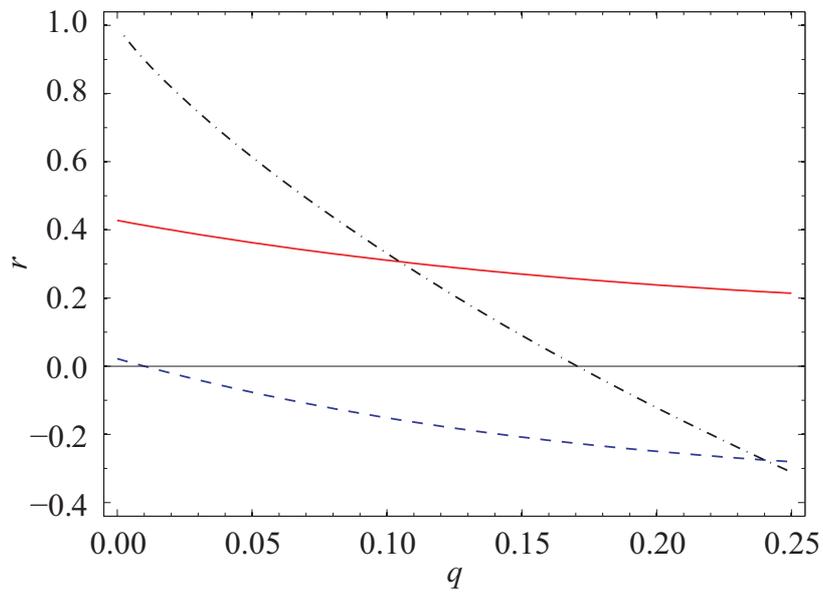}
\caption{Relative key rate as a function of the QBER in BB84. Quantum channel noise is given by the operator $N_{xyz}(n)$, and qubit loss is fixed to 50\%. The rate for gate $U_2^{\displaystyle{\star}}$ assuming full correlation of errors in qubit pairs is shown in red solid line and without any correlation of errors in blue dashed line. The BB84 scheme, without any correlation of errors, is shown in black dash-dotted line. The BB84 scheme yields a higher key rate for approximately $q < 10.5\%$.}
\label{fig:q-corrate}
\end{center}
\end{figure}

Correlation of errors is not a reasonable assumption with BB84, either. Let us therefore compare the relative key rate of Eq.~(\ref{eq:corrrate}) for gate $U_2^{\displaystyle{\star}}$, which is expected to provide at least some correlation of errors, with the rate of Eq.~(\ref{eq:ratebb84}) for the BB84 scheme. Figure \ref{fig:q-corrate} shows these rates as a function of $q$, assuming quantum channel noise given by Eq.~(\ref{eq:noisexyz}) and a 50\% qubit loss. Gate $U_2^{\displaystyle{\star}}$ assuming full correlation of errors is shown in red solid line, while BB84 assuming no correlation is shown in black dash-dotted line. The former yields a higher key rate for approximately $q > 10.5\%$. The rate also seems to remain positive when the QBER increases. The blue dashed line shows the rate for $U_2^{\displaystyle{\star}}$ without any correlation of errors. This rate is negative for approximately $q > 1\%$.

%include{reeqkd} jos ehtii (toissijainen) (tulokset hyvin laihoja)
\chapter{Entanglement-Enhanced Attacks}
\label{ch:eeattack}

This chapter presents the entanglement-enhanced (EE) attack against the EEQKD protocol for two- and three-qubit entanglement. The attack is based on Eve locally manipulating entangled qubits, one or more of which are already in Bob's possession. For the gates $U_2^{\displaystyle{\star}}$ and $U_3^{\displaystyle{\star}}$, defined in Sec.~\ref{sec:ustar}, the EE attack reduces the EEQKD protocol to BB84. Here, Eve utilizes $N$-qubit entanglement to fully counteract the restriction posed by Alice and Bob's use of $N$-qubit entanglement. The two-qubit EE attack was partially developed in Ref.~\cite{dippa}.

The technological requirements for the attack are a subset of those for Alice and Bob. Eve needs to utilize short-term quantum memory, perform projective measurements as well as quantum non-demolition measurements to detect the presence of an intercepted qubit. In addition, Eve needs to apply single-qubit gates chosen from a small set of gates, and she needs to be able to repeatedly construct a constant entangled state of two or three qubits. The latter requirements translate to the application of a local $N$-qubit gate and to applying an entangling $N$-qubit gate to a constant state.

\section{Two-Qubit Protocol}

For clarity of presentation and without loss of generality, we assume that Alice and Bob have chosen $U_{2,\mathrm{even}}^{\displaystyle{\star}+}$ as the gate $U_2^{\displaystyle{\star}}$. The superscript '$+$' denotes the sign chosen in Eq.~(\ref{stargate}). Changing the parity or the sign of $U_2^{\displaystyle{\star}}$ is accommodated by straightforward changes of sign and bit value in Eve's actions. The states transmitted by Alice are
\begin{equation}
U_2^{\displaystyle{\star}} |a_1;\alpha_1\rangle |a_2;\alpha_2\rangle = \left[ |a_1 a_2\rangle \pm (-1)^{a_1+a_2} i |\bar{a}_1 \bar{a}_2\rangle \right] / \sqrt{2},
\label{eq:u2starstates}
\end{equation}
where the upper sign is chosen if and only if $\alpha_1 \neq \alpha_2$. The bases $\alpha_1$ and $\alpha_2$ are unchanged, and omitted to enhance readability on the right-hand side of Eq.~(\ref{eq:u2starstates}). Furthermore, we have employed the notation $|\psi\rangle \otimes |\varphi\rangle = |\psi\rangle |\varphi\rangle = |\psi \varphi\rangle$.

\subsection{Obtaining the Qubit Pair}
To begin the attack, Eve captures the first, i.e., left-slot, qubit of the state in Eq.~(\ref{eq:u2starstates}) and stores it in quantum memory. She then sends the left-slot qubit of the EPR state
\begin{equation}
|\mathrm{Eve}_2^0\rangle = \left( |0\rangle |0\rangle - i |1\rangle |1\rangle \right) / \sqrt{2}
\label{eq:eve02}
\end{equation}
to Bob, who acknowledges receiving it, thus triggering Alice to send the second, right-slot, qubit of the state in Eq.~(\ref{eq:u2starstates}) to Eve. Having received both qubits, Eve applies $U_2^{\displaystyle{\star} \dagger}$ to the qubit pair, and is now in possession of the state $|a_1;\alpha_1\rangle |a_2;\alpha_2\rangle$.

\subsection{Measurement and Entangled-State Reconstruction}
Eve measures the product-state qubits in the $\sigma_z$ eigenbasis. Depending on the individual results $e_1$ and $e_2$ from the left- and right-slot qubits, respectively, Eve applies the single-qubit operator
\begin{equation}
E_{e_1e_2} = (-i \sigma_y)^{e_2} (i \sigma_x)^{e_1},
\end{equation}
to the right-slot qubit of Eq.~(\ref{eq:eve02}). She then sends this qubit to Bob, who has the state
\begin{equation}
\left( I_1 \otimes E_{e_1e_2} \right) |\mathrm{Eve}_2^0\rangle = \left( |e_1 e_2\rangle - (-1)^{e_1+e_2} i |\bar{e}_1 \bar{e}_2\rangle \right) / \sqrt{2}.
\end{equation}
That is, Eve is able to send the state $U_2^{\displaystyle{\star}} |e_1;z\rangle |e_2;z\rangle$ to Bob.

\subsection{Mutual Information and QBER}
\label{subsec:mi-qber}
Bob applies $U_2^{\displaystyle{\star} \dagger}$ to the received entangled qubit pair, hence obtaining the state $|e_1;z\rangle |e_2;z\rangle$. This shows that the EEQKD protocol has been reduced to BB84, and the effect of the attack is the same as an IR attack in BB84. With probability $\frac{1}{4}$, the bases $\alpha_1 = \alpha_2 = z$ and Eve gains full information on the bit values while inducing no errors. With probability $\frac{1}{2}$, exactly one of the two bases is $z$ and Eve gains 0.5 information and induces a 25\% QBER, on average. With probability $\frac{1}{4}$, the bases $\alpha_1 = \alpha_2 = x$, in which case Eve gains no information and induces an average QBER of 50\%. The expected values of information gain and QBER are thus 0.5 and 25\%, respectively.

\section{Three-Qubit Protocol}

For clarity of presentation and without loss of generality, we again assume that Alice and Bob have chosen $U_{3,\mathrm{even}}^{\displaystyle{\star}+}$ as the gate $U_3^{\displaystyle{\star}}$. Changing the parity or the signs of $U_3^{\displaystyle{\star}}$ is accommodated by straightforward changes of sign and bit value in Eve's actions. In this case, Alice transmits states
\begin{multline}
U_3^{\displaystyle{\star}} |a_1;\alpha_1\rangle |a_2;\alpha_2\rangle |a_3;\alpha_3\rangle =  \\
\frac{1}{2} \left[ |a_1 a_2 a_3\rangle \pm (-1)^{a_2+a_3} i |a_1 \bar{a}_2 \bar{a}_3\rangle \pm (-1)^{a_1+a_2} i |\bar{a}_1 \bar{a}_2 a_3\rangle \mp (-1)^{a_1+a_3} |\bar{a}_1 a_2 \bar{a}_3\rangle \right],
\label{eq:u3starstates}
\end{multline}
where the upper sign is chosen for the second term iff $\alpha_2 \neq \alpha_3$, for the third term iff $\alpha_1 \neq \alpha_2$, and for the fourth term iff $\alpha_1 \neq \alpha_3$. The bases $\alpha_1$, $\alpha_2$, and $\alpha_3$ are unchanged, and again omitted to enhance readability on the right-hand side of Eq.~(\ref{eq:u3starstates}).

\subsection{Processing the First Two Qubits}

To begin the attack, Eve captures the left-slot qubit of the state in Eq.~(\ref{eq:u3starstates}) and stores it in quantum memory. She then sends the left-slot qubit of the state
\begin{equation}
|\mathrm{Eve}_3^0\rangle = \frac{1}{2} \left( |0\rangle |0\rangle |0\rangle - i |0\rangle |1\rangle |1\rangle - i |1\rangle |1\rangle |0\rangle + |1\rangle |0\rangle |1\rangle \right),
\label{eq:eve03}
\end{equation}
to Bob, who acknowledges receiving it, thus triggering Alice to send the second, middle-slot, qubit of the state in Eq.~(\ref{eq:u3starstates}) to Eve. Having received the first two qubits, Eve applies to them the operation $-(\sigma_y \otimes I_1) U_{2,\mathrm{even}}^{\displaystyle{\star}+}$. The full operation applied to the initial product state $|a_1;\alpha_1\rangle |a_2;\alpha_2\rangle |a_3;\alpha_3\rangle$ so far is
\begin{equation}
\left\{ \left[ -(\sigma_y \otimes I_1) U_{2,\mathrm{even}}^{\displaystyle{\star}} \right] \otimes I_1 \right\} U_3^{\displaystyle{\star}} = \left( I_1 \otimes I_1 \otimes \sigma_y - i I_1 \otimes \sigma_y \otimes I_1 \right) /\sqrt{2} = I_1 \otimes U_{2,\mathrm{odd}}^{\displaystyle{\star}-}.
\label{eq:first2op}
\end{equation}
Therefore, the first, left-slot, qubit of the transmitted state is unentangled from the other two. This enables Eve to measure the first qubit in the $z$ basis without interfering the last two qubits. Eve stores the received middle-slot qubit in quantum memory. If and only if the result $e_1$ of the first measurement is 1, Eve applies $\sigma_z$ to the middle-slot qubit of Eq.~(\ref{eq:eve03}).  She then sends the qubit to Bob.

\subsection{Processing the Last Two Qubits}

Bob's reception of the second, middle-slot qubit, triggers Alice to send the last, right-slot qubit to Eve. As shown in Eq.~(\ref{eq:first2op}), the last two qubits are now in state $U_{2,\mathrm{odd}}^{\displaystyle{\star}-} |a_2;\alpha_2\rangle |a_3;\alpha_3\rangle$. Eve applies $U_{2,\mathrm{odd}}^{\displaystyle{\star}- \dagger}$ to this state, and measures the qubits $|a_2;\alpha_2\rangle$ and $|a_3;\alpha_3\rangle$ in the $z$ basis. Let $e_2$ and $e_3$ denote the measurement results, respectively.

The single-qubit gate to be applied to the last qubit of Eq.~(\ref{eq:eve03}) depends on all three results $e_1$, $e_2$, and $e_3$. If $e_1 = 0$, the operation is equal to that of the two-qubit case, $(-i \sigma_y)^{e_2} (i \sigma_x)^{e_1}$. If $e_1 = 1$, the operation is $i \sigma_y (-i \sigma_y)^{e_3} (-i \sigma_x)^{e_2}$. Hence, the full operation applied to the state $|\mathrm{Eve}_3^0\rangle$ is
\begin{equation}
E_{e_1 e_2 e_3} = I_1 \otimes (\sigma_z)^{e_1} \otimes \left[ \delta_{e_1,0} (-i \sigma_y)^{e_3} (i \sigma_x)^{e_2} + \delta_{e_1,1} i \sigma_y (-i \sigma_y)^{e_3} (-i \sigma_x)^{e_2} \right],
\end{equation}
where $\delta_{r,v}$ is the Kronecker delta. These single-qubit operations allow Eve to send the state $U_3^{\displaystyle{\star}} |e_1;z\rangle |e_2;z\rangle |e_3;z\rangle$ to Bob.

\subsection{Mutual Information and QBER}
Bob applies $U_3^{\displaystyle{\star} \dagger}$ to the received entangled qubit triplet, and thus obtains the state $|e_1;z\rangle |e_2;z\rangle |e_3;z\rangle$. This shows that the EEQKD protocol has been reduced to BB84, and the effect of the attack is the same as an IR attack in BB84. The calculation of Eve's mutual information and the induced QBER is similar to that presented in Sec.~\ref{subsec:mi-qber}. On average, Eve gains 0.5 bits of information per key bit and induces an average QBER of 25\%.

\chapter{Conclusions}
\label{ch:conclusions}

This Licentiate Thesis presentes an in-depth investigation of the EEQKD protocol, which is based on transmitting $N$-qubit entangled states. It is shown that the protocol cannot be less secure than the well-known BB84 protocol, on which EEQKD is based. We have shown that by entangling the qubits optimally, it is possible to significantly reduce the maximal information provided by an intercept-resend attack for a given error rate. We have presented an $N$-qubit gate, $U_N^{\displaystyle{\star}}$, which restricts the information gain of an IR attack to at most $1/(2N)$.

Chapter \ref{ch:keyrate} discusses the arguably most important figure for a QKD protocol, the key generation rate, in the case $N=2$. A model is developed for the inherent noise in the quantum channel, and the effect of this noise to the key rate is estimated. The loss of qubits in the quantum channel also strongly affects the key rate. The effect is estimated for a 50\% loss, roughly corresponding to a transmission distance of 10km. It is shown that at least for gate parameter values near $\mathbf{c} = (0,\frac{\pi}{2},0)$, i.e., gate $U_2^{\displaystyle{\star}}$, the combined effect of noise and loss makes the error rate prohibitively large.

The relative key rate was maximized over the entangling gate $U_2$ for various values of the qubit error rate, taking into account quantum channel noise. Without qubit loss, the gate $C \left( \frac{\pi}{4},\frac{\pi}{4},\frac{\pi}{4} \right)$ allows for fastest key generation. The key rate is between 3\% to 35\% higher than that of BB84 for typical values of the qubit error rate, and the difference increases with the error rate. Errors or loss introduced by the additional equipment of the EEQKD protocol were not taken into account. An extensive effort was made to find a gate $U_2$ yielding a higher key rate than BB84 with noise as well as qubit loss, but no such gate was discovered. This study conclusively shows that in present day practical applications, BB84 performs better than the EEQKD protocol---increasing entanglement between consecutive qubits decreases the key rate.

For highly entangled qubit pairs, the bit errors may be correlated. If this is taken into account in the protocol, it may reduce the information leaked to an attacker in the error correction phase. For $N=2$, the correlation ideally halves the leaked information. It is shown in Sec.~\ref{sec:errcorr}, that gate $U_2^{\displaystyle{\star}}$ could yield a higher key rate than BB84 for error rate values larger than 10.5\%, if full correlation of errors is assumed. This is in spite of the detrimental effects of quantum channel imperfections in particular for the gate $U_2^{\displaystyle{\star}}$.

A novel attack against the EEQKD protocol was developed based on an earlier partial presentation. An attacker is able to reduce the EEQKD protocol to the BB84 protocol at least for the gates $U_2^{\displaystyle{\star}}$ and $U_3^{\displaystyle{\star}}$. It is expected that the attack generalizes to an arbitrary value of $N$, but for values of the gate parameter $c_2$ between 0 and $\frac{\pi}{2}$ the generalization is unlikely. This is because the absolute values of the amplitudes of the superposition states are unequal for these gates, and the amplitudes cannot be altered with the approach discussed. However, the attack may provide more information than an IR attack for other gates, as well. It is not surprising that the key to Eve's overcoming of the inaccessibility to the full multi-qubit entangled state lies in the equally clever use of entangled qubits.

Although the results in Ch.~\ref{ch:infoqber} strongly suggest that the protocol performs better than the BB84 scheme, the studies regarding the key generation rate, in Ch.~\ref{ch:keyrate}, in fact show that the opposite may be true. In practical key distribution applications at present, one cannot recommend the use of EEQKD over BB84. In the scenario of very low qubit loss, however, the EEQKD protocol may perform better, especially if the qubit error rate is relatively high. Furthermore, the attack described in Ch.~\ref{ch:eeattack} allows Eve to reduce EEQKD with $U_N^{\displaystyle{\star}}$ to the plain BB84 scheme, at least in the case of two or three qubit entanglement. This raises the question whether the attack can be modified to overcome any entangling gate.

Correlation of bit errors in entangled qubits warrants further study. High correlation of errors in an $N$-qubit group would significantly reduce the information leaked in error correction, which is the main contributor to the decrease in the key rate of the EEQKD protocol. Unless full correlation is achieved, the protocol needs to be amended to take the correlation into account properly.

We have shown that full correlation of bit errors could result in a key rate above that of the BB84 scheme, if the underlying error rate is high. This, however, only applies to a fairly low value of qubit loss, 50\%. In a realistic scenario, the loss is of the order of 0.999 for transmission distances beyond 100km. Therefore, developing states robust to qubit loss is a topic for future research. The difficulty here lies in balancing the need to protect the transmission from an attacker with the need to tolerate uncontrolled loss of qubits.

In the course of this study, it has become manifest that it is usually beneficial for an attacker to measure all the qubits of an entangled state, as opposed to measuring only a fraction of qubits in each group. Hence, interleaving and randomizing the qubits of different entangled groups for the transmission could improve the information-error ratio. This is yet another topic for future research.

\appendix
\chapter{List of Source Code Files}
\label{ch:codefiles}

%lisää reeqkd-filet jos siitä tulee tekstiä työhön

This appendix lists and briefly delineates the files containing the source code used to perform the calculations presented in this thesis. A more detailed description is given in each file. File names ending with '.m' are scripts or functions for Matlab by The MathWorks, Inc., and those ending with '.nb' are notebooks for Mathematica by Wolfram Research, Inc. 

\begin{description}

\item[init2.m] initializes the two-qubit BB84 states in the array \texttt{msg}.

\item[applySU2full.m] applies a single-qubit gate (SU2 matrix) to one of the qubits of each pair in the array \texttt{msg}.
\item[A.m] applies gate $C(\mathbf{c})$ to each of the states in the array \texttt{msg}.
\item[applySU2.m] applies successive $z$ and $y$ rotations to one of the qubits of each pair in the array \texttt{msg}.
\item[measure1.m] gives the single-qubit projective measurement operator for the given result and basis.
\item[measure2.m] gives the two-qubit projective measurement operator for the given results and bases.
\item[summsgs.m] sums over different values of $a$ in the probability matrix of Bob's results for given $\alpha$ and $e$.
%are all that the below one needs
\item[bb84e34.m] calculates the mutual information and QBER for given gate parameters for Alice and Bob's and Eve's gates.

\item[slopebb84e34.m] returns the slope $I(A,E)/q$ for the given gate parameters for Alice and Bob and for Eve.
%is used by
\item[bb84emaxeve34.m] maximizes the slope 9 times over Eve's parameters and returns the maximal value found.
%is used by
\item[bb84eminmax34.m] minimizes the slope over Alice and Bob's parameters and returns the minimal slope and the corresponding parameters.

\item[A3.m] applies gate $C(\mathbf{c})$ to the two rightmost states in each three-qubit state given as a parameter.
%is used by
\item[bb84e40q.m] simulates the protocol under photon loss. The first entangled photon is replaced with the given state. Returns only the QBER's of the two qubits and their average.

\item[applynoise.m] applies the noise operator $N_{xyz}$ to the specified qubit in each of the pairs in the array \texttt{msg}.
%is used by
\item[bb84e60.m] simulates the protocol under innocent quantum channel noise with a given amplitude. Returns only the QBER's of the two qubits and their average.

\item[Hbin.m] gives the binary Shannon entropy.
%is used by
\item[eeqkdrate4.m] gives the relative key rate according to Eq.~(\ref{eq:relkeyrate}) for the given gate parameters, $q$, and noise amplitude. Assumes 50\% qubit loss. The slope $s$ is maximized.
%is used by
\item[eeqkdmaxmaxkeyrate4.m] maximizes the relative key rate for the given noise amplitude over the parameters $\mathbf{c}$.
%is used by
\item[optimrate4.m] optimizes the relative key rate for different BB84 QBER's.

\item[qber.nb] contains calculations and plots regarding the key rate with noise and loss.
\item[secret-key\_rate.nb] includes commands to produce the plots in Figs.~\ref{fig:r-delta} and \ref{fig:u2star-n-rates}.
\item[eprattack.nb] contains calculations needed in the analysis of the entanglement-enhanced attacks discussed in Ch.~\ref{ch:eeattack}.

\end{description}

\setlength{\baselineskip}{1ex}

%\addtolength{\textheight}{2.5cm} (muuten sivu jatkuu liian alas)
\addcontentsline{toc}{chapter}{Bibliography}
\bibliographystyle{mysty}
 %{\footnotesize{\bibliography{BEC,MyBec}}}
\bibliography{lis}
%\addcontentsline{toc}{chapter}{Index}	
%\printindex	
\end{document}